\documentclass{appolb}
\usepackage{graphicx,color}

\begin{document}
\title{Light cluster production at NICA%
\thanks{Presented at Critical Point and Onset of Deconfinement 2016, Wroc{\l}aw}%
}
\author{N.-U. Bastian$^{a}$, D. Blaschke$^{a,b,c}$, G. R\"opke$^{c,d}$
\address{$^{a}$Institute of Theoretical Physics, University of Wroclaw, Wroclaw, Poland
\\$^{b}$Bogoliubov Laboratory for Theoretical Physics, JINR Dubna, Dubna, Russia
\\$^{c}$National Research Nuclear University, MEPhI, Moscow, Russia
\\$^{d}$University of Rostock, D-18051 Rostock, Germany}
}
\maketitle

\begin{abstract}
In a recent publication \cite{Bastian:2016xna} the importance of measuaring light clusters at future NICA and FAIR experiments as ``rare probes'' for in-medium characteristics was presented.
Within theoretical conciderations is shown how clusters would be affected by medium affects as given collider energies.  
\end{abstract}


\section*{}
To decide on characteristics of the matter produced in energetic nuclear collisions a variety of probes is needed that provide different perspectives.
One class of such probes are nuclear clusters.
Here we discuss the insights that the clusters can provide within the NICA program, both within fixed 
target and collider setups, at BM@N and MPD, respectively. 
 
\begin{figure}[!t]
\resizebox{\textwidth}{!}{%
\includegraphics{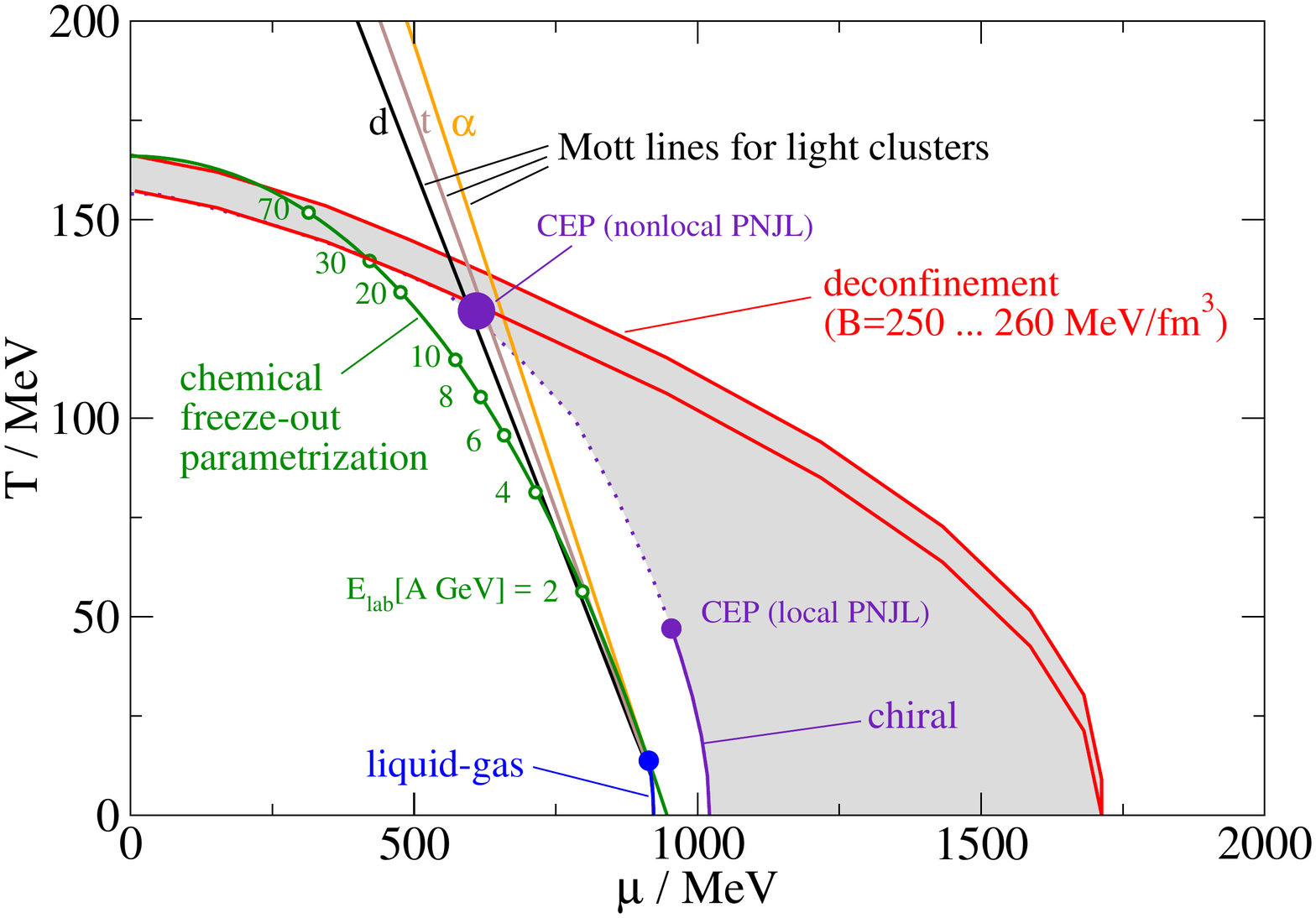}
}
\caption{Phase diagram of dense nuclear matter in the plane of temperature $T$ and baryochemical potential $\mu_B$.
The diagram includes Mott-lines for the dissociation of light nuclear clusters, extrapolated also to the deconfinement region. For details, see text.
\label{PD-Mott}
}
\end{figure}

To provide a general background we show in Fig.~\ref{PD-Mott} a phase diagram of dense nuclear matter including lines for Mott dissociation of deuterons ($d$), tritons ($t$) and alpha particles 
($\alpha$), taken from \cite{Schuck:1999eg}, see also \cite{Ropke:1983lbc},  together with the parametrization of the chemical freeze-out line \cite{Begun:2012rf} from statistical model fits of hadron production in heavy-ion collisions. 
Several laboratory energies from the energy range accessible at the 
NICA accelerator complex are shown as labeled dots on that line.

For the sake of generating the illustration, the hadronic phase is described by a DD2 equation of state 
\cite{Typel:2009sy}
with a liquid-gas phase transition (blue line with critical endpoint), extended by adding components of the hadron resonance gas, in particular pions and kaons.
The quark-gluon matter phase for the figure is described by a PNJL model exhibiting a first order/crossover chiral transition with a critical endpoint position that can be adjusted with details of the model (local or nonlocal, see \cite{Contrera:2016rqj}). 
The deconfinement lines are obtained requiring the baryon chemical potential and pressure in the DD2 hadronic and PNJL quark matter to be equal. 
The latter pressure is shifted by a (gluonic) bag constant of either 250 or 260 MeV/fm$^3$, respectively.

If the freeze-out model, describing hadron abundances in terms of specific temperature and chemical potential values, is extended to nuclear clusters, the law of mass action (LMA) \cite{DasGupta:1981xx} 
applies to cluster abundancies.
Under LHC conditions, the freeze-out model 
\cite{Andronic:2010qu,Cleymans:2011pe,Andronic:2011yq}
appears to describe the abundancies very accurately
\cite{Adam:2015vda}.
Very recently, the data for production of light nuclei in the NA49 experiment at SPS have been published which are interpreted using the statistical model too \cite{Anticic:2016ckv}.  
  
In terms of laboratory energy per nucleon $E_{\rm lab}$, for fixed target, the freeze-out curve can be approximated with \cite{Begun:2012rf}

\subsection*{Acknowledgement}
N.-U.B., D.B. acknowledge support by the Polish NCN under grant number UMO-2011/02/A/ST2/00306.
D.B. acknowledges support by the MEPhI Academic Excellence Project under contract No. 02.a03.21.0005.

{}


\begin{thebibliography}{}

\bibitem{Bastian:2016xna}
  N.-U.~Bastian {\it et al.},
  Eur.\ Phys.\ J.\ A {\bf 52} (2016) no.8,  244

\bibitem{Schuck:1999eg} 
  P.~Schuck {\it et al.},
  Few Body Syst.\ Suppl.\  {\bf 12}, 50 (2000).

\bibitem{Ropke:1983lbc}
  G.~R\"opke {\it et al.},
  Nucl.\ Phys.\ A {\bf 399} (1983) 587.
 
\bibitem{Begun:2012rf} 
  V.~V.~Begun {\it et al.},
  Phys.\ Rev.\ C {\bf 88}, no. 2, 024902 (2013)

\bibitem{Typel:2009sy} 
  S.~Typel {\it et al.},
  Phys.\ Rev.\ C {\bf 81}, 015803 (2010).

\bibitem{Contrera:2016rqj}
  G.~A.~Contrera {\it et al.},
  Eur.\ Phys.\ J.\ A {\bf 52} (2016) no.8,  231

\bibitem{DasGupta:1981xx} 
  S.~Das Gupta {\it et al.},
  Phys.\ Rept.\  {\bf 72}, 131 (1981).

\bibitem{Andronic:2010qu} 
  A.~Andronic {\it et al.},
  Phys.\ Lett.\ B {\bf 697}, 203 (2011).

\bibitem{Cleymans:2011pe} 
  J.~Cleymans {\it et al.}
  Phys.\ Rev.\ C {\bf 84}, 054916 (2011).

\bibitem{Andronic:2011yq} 
  A.~Andronic {\it et al.},
  J.\ Phys.\ G {\bf 38}, 124081 (2011).


\bibitem{Adam:2015vda} 
  J.~Adam {\it et al.} [ALICE Collaboration],
  Phys.\ Rev.\ C {\bf 93}, 024917 (2016).

\bibitem{Anticic:2016ckv} 
  T.~Anticic {\it et al.} [NA49 Collaboration],
  arXiv:1606.04234 [nucl-ex].

\end{thebibliography}
\end{document}